\documentclass[journal=jacsat,manuscript=article]{achemso}
\usepackage[utf8]{inputenc}
\usepackage[version=3]{mhchem} % Formula subscripts using \ce{}
\usepackage{textgreek}

\usepackage{graphicx, wrapfig, lipsum}
\usepackage{achemso}
\usepackage[format=plain,justification=justified,singlelinecheck=false,labelfont=bf,labelsep=space]{caption}
\usepackage{subcaption}
\usepackage{todonotes}
\usepackage{float}
\usepackage{amsmath}
\usepackage{url}
\usepackage{amsmath}
\usepackage{amsfonts}
\usepackage{hyperref}
\usepackage[ruled,vlined]{algorithm2e}

\usepackage{setspace}
\usepackage{xr}
\newcommand{\hmn}[1]{% Hermann-Maguin notation
  \ensuremath{\begingroup\setupHMN #1\endgroup}%
}

\newcommand{\setupHMN}{%
  \doHMN{-}{\HMNoverline}%
  \doHMN{*}{\HMNminverse}%
  \doHMN{i}{\infty}
}

\newcommand{\doHMN}[2]{%
  \begingroup\lccode`~=`#1
  \lowercase{\endgroup\let~}#2%
  \mathcode`#1="8000
}

\newcommand{\HMNminverse}[1]{\frac{#1}{m}}
\newcommand{\HMNoverline}[1]{\mkern1mu\overline{\mkern-1mu#1\mkern-1mu}\mkern1mu}

\author{Yuxing Fei}
\altaffiliation{Authors contribute equally}
\affiliation[ucb]{Department of Materials Science \& Engineering, University of California, Berkeley, CA 94720, USA}
\alsoaffiliation[lbl]{Materials Sciences Division, Lawrence Berkeley National Laboratory, Berkeley, CA 94720, USA}

\author{Matthew J. McDermott}
\altaffiliation{Authors contribute equally}
\affiliation[lbl]{Materials Sciences Division, Lawrence Berkeley National Laboratory, Berkeley, CA 94720, USA}

\author{Christopher L. Rom}
\affiliation[nrel]{Materials Science Center, National Renewable Energy Laboratory, Golden, CO, 80401, USA}

\author{Shilong Wang}
\affiliation[ucb]{Department of Materials Science \& Engineering, University of California, Berkeley, CA 94720, USA}
\alsoaffiliation[lbl]{Materials Sciences Division, Lawrence Berkeley National Laboratory, Berkeley, CA 94720, USA}

\author{Gerbrand Ceder}
\affiliation[ucb]{Department of Materials Science \& Engineering, University of California, Berkeley, CA 94720, USA}
\alsoaffiliation[lbl]{Materials Sciences Division, Lawrence Berkeley National Laboratory, Berkeley, CA 94720, USA}
\email{gceder@berkeley.edu}

\title{Dara: Automated multiple-hypothesis phase identification and refinement from powder X-ray diffraction}

\begin{document}
\maketitle
\newpage
\begin{abstract}
Powder X-ray diffraction (XRD) is a foundational technique for characterizing crystalline materials. However, the reliable interpretation of XRD patterns, particularly in multiphase systems, remains a manual and expertise-demanding task. As a characterization method that only provides structural information, multiple reference phases can often be fit to a single pattern, leading to potential misinterpretation when alternative solutions are overlooked. To ease humans' efforts and address the challenge, we introduce Dara (Data-driven Automated Rietveld Analysis), a framework designed to automate the robust identification and refinement of multiple phases from powder XRD data. Dara performs an exhaustive tree search over all plausible phase combinations within a given chemical space and validates each hypothesis using a robust Rietveld refinement routine (BGMN). Key features include structural database filtering, automatic clustering of isostructural phases during tree expansion, peak-matching-based scoring to identify promising phases for refinement. When ambiguity exists, Dara generates multiple hypothesis which can then be decided between by human experts or with further characteriztion tools. By enhancing the reliability and accuracy of phase identification, Dara enables scalable analysis of realistic complex XRD patterns and provides a foundation for integration into multimodal characterization workflows, moving toward fully self-driving materials discovery.
\end{abstract}
\newpage
\section{Introduction}
% XRD is important. XRD analysis was performed to obtain valuable information
Accurate characterization of material structures is essential for understanding synthesis-structure-property relationships in materials. For bulk inorganic materials, powder X-ray diffraction (XRD) has long been a pivotal and widely used technique for determining majority-phase crystal structures \cite{waseda2011x, ward2017automated}. Moreover, powder XRD plays a critical role in the discovery of inorganic materials, serving as a key tool for confirming the synthesis of predicted target structures, such as those derived from \textit{ab initio} methods \cite{narayan2016computational, shoemaker2014situ}. With appropriate analysis, XRD patterns can provide valuable insights into material properties, including phase fractions, lattice parameters, strains, site occupancies, and more \cite{mccusker1999rietveld}.

% How to perform XRD analysis
% phase identification
XRD analysis typically begins with the identification of all phases present in the pattern. This process involves comparing the experimental pattern with the calculated patterns of structures available in crystal structure databases like the Materials Project (MP) \cite{jain2013commentary}, Inorganic Crystal Structure Database (ICSD) \cite{zagorac2019recent}, Powder Diffraction File (PDF) \cite{kabekkodu20245}, and Crystallography Open Database (COD) \cite{Grazulis2012}. The task becomes practically challenging when a sample contains multiple phases that cannot be perfectly matched to reference structures. This is common when characterizing the synthesis products of exploratory inorganic reactions or natural minerals, which may exhibit compositional variance or preferred orientation effects.
% Human expert's insights are necessary.
In these cases, accurate interpretation requires meticulous analysis and the expertise of researchers who are intimately familiar with the material system, enabling them to discern its subtle nuances.

% XRD needs human involvement
% However, established as the XRD analysis method is, to extract as much information as possible from XRD patterns, human experts are usually needed in the process, with a lot of caution taken to ensure a valid interpretation of the spectra. This is because XRD is a structure-based technique and inherently loses information about a sample's composition. Therefore, a diffraction pattern can sometimes be interpreted as many isostructural phases with varied compositions. \cite{leeman2024challenges, holder2019tutorial}. Distinguishing the nuances between phases sometimes requires knowledge about the sample's preparation procedure, chemistry, and synthesis condition, which can be found nowhere in an XRD pattern. 

% Automated XRD analysis is important.
In recent years, the development of automated and autonomous laboratories for the discovery of inorganic materials has underscored the need to accelerate the characterization process \cite{szymanski2023autonomous, lunt2024modular, chen2024navigating, yotsumoto2024autonomous}. As the throughput of synthesis and characterization continues to increase, human analysis of patterns has become impractical, further emphasizing the importance of automation. The integration of reliable, robust, and accurate powder XRD phase identification algorithms within autonomous laboratories will be crucial to establishing high-quality experimental databases for inorganic materials. In self-driving labs, high-quality characterization of samples at the early stage of synthesis optimization is of particular importance in the AI-driven decision-making algorithms, which may be more challenged by erroneous interpretations than human experts are.

% Current phase identification software
% Many recent works have focused on establishing accurate automated phase identification algorithms based on XRD pattern analysis.
Algorithms for XRD phase analysis have captured the interest of researchers for nearly a century. In 1938, Hanawalt \cite{hanawalt1938chemical} introduced a qualitative method for phase identification based on major diffraction peaks of known phases. Using this approach, researchers can refer to the ``Hanawalt Index'', a tabulated collection of peak data, to identify potential phases. If three major peaks align with those of a specific phase, it strongly suggests the presence of that phase in the sample. This method is also known as the ``search-match'' method, as it always involves searching the reference database and then matching the phase to the diffraction pattern. Following Hanawalt's manual search, several computer programs are available for automated peak indexing \cite{smith1991powder, lin1993novel, dinnebier2015powder}, each with carefully fine-tuned strategies to identify phases. With more peaks taken into consideration and user-friendly interfaces, these programs enable researchers to analyze XRD patterns with rigor and ease. This makes them the dominant method for phase analysis due to their straightforward nature and low computational resource requirements.

% NN-based methods
With the growing trend of applying deep learning to XRD analysis, numerous studies have leveraged neural networks (NNs) to automate the interpretation of diffraction patterns. Oviedo et al. \cite{oviedo2019fast} used a convolutional neural network (CNN) with simulated and experimental XRD data to classify crystal dimensionality and space group, achieving 93\% and 89\% accuracy, respectively, by augmenting limited data with physics-informed simulations. Lee et al.\cite{lee2020deep} tackled multi-phase identification by training a deep CNN on over 1.7 million synthetic mixed XRD patterns (combinations of 170 compounds in the Sr-Li-Al-O quaternary system), enabling near-100\% phase identification in complex mixtures and even quantifying phase fractions with 86\% accuracy. Maffettone et al. \cite{maffettone2021crystallography} employed an ensemble of CNNs (a “crystallography companion agent”) that outputs probabilistic phase predictions, avoiding combinatorial explosion in training while providing confidence metrics for each identified phase. Szymanski et al.\cite{szymanski2021probabilistic} combined physics-informed peak perturbation augmentation with an ensemble CNN and a branching algorithm to iteratively identify phases in mixtures. This probabilistic approach achieved higher accuracy than traditional profile-matching and earlier deep-learning methods on challenging multi-phase samples. More recently, researchers have created a new NN architecture for the XRD phase identification task. For example, Zhang et al. \cite{zhang2024crystallographic} introduced a self-attention CNN (CPICANN) trained on around 700k simulated patterns (from 23k structures), which attained 98.5\% accuracy on single-phase identification with element information provided and 80\% on experimental scans, significantly outperforming conventional XRD software. Beyond deep NNs, other machine-learning approaches have also been explored. For example, Suzuki et al. \cite{suzuki2020symmetry} used an interpretable tree-ensemble model to classify crystal systems and space groups with 90\% accuracy, revealing human-understandable diffraction features. 

% Refinement-based method
Alongside computerized Hanawalt methods and their NN-based variations, full-profile search-match methods have recently gained attention. These methods use pattern-fitting programs to match calculated patterns to experimental data. For example, Lutterotti et al.\cite{lutterotti2019full} used results from Rietveld refinement to calculate a figure of merit (FoM) that measures phase fitness, considering factors such as the refinement R-value, density differences, crystallite size, and microstrain. Chang et al.\cite{chang2023probabilistic} proposed a pseudo-refinement method called CrystalShift, which uses a best-first tree search to refine phase combinations and applies Bayesian model comparison to estimate phase probabilities without requiring additional phase space information or training. These methods can use more detailed peak models to handle patterns with poor crystallinity or highly oriented grains. They also improve interpretability by providing additional information through the refinement process, helping to better understand the fitness of the output phase combination.

% XRD is only a structural characterization tool
Despite the numerous approaches proposed for reliable phase identification in XRD, they still occasionally yield inaccurate results. This limitation arises from the nature of XRD as a structure-based technique, which inherently lacks information about the composition of phases \cite{leeman2024challenges, holder2019tutorial}. A diffraction pattern can often be interpreted as different combinations of reference phases due to the presence of isostructural phases and solid solutions in the structure database. Furthermore, when a sample contains a mixture of phases, the minor peaks of some phases may overlap with the dominant peaks of others. This peak overlap can result in several plausible phase combinations that fit the pattern, making it impossible to definitively determine the correct solution based solely on XRD data without external knowledge of the material system.

% What does the XRD phase identification software need?
Given these challenges, a reliable automated XRD analysis tool should be able to present multiple possible phase combinations matching a pattern when ambiguity exists. For example, when characterizing a synthesized sample of a purported solid solution, the tool should be able to present and compare the null hypothesis (a multi-phase combination of end-members) with the desired result (a single-phase solid solution). This is especially critical whenever the endmembers are isostructural with the target solid solution. Moreover, in cases where the null or alternative hypotheses have a similar quality of fit, the tool should be able to provide hints for further characterization to disambiguate different solutions.

% \todo{Update this paragraph. Make it longer and more detailed}
% Introduce Dara
Motivated by the need to address the ambiguity issue in automated XRD pattern analysis, we present Dara, a data-driven Rietveld analysis framework. Dara is designed to generate all validated hypotheses that align well with a given XRD pattern, offering a comprehensive and reliable solution for phase identification.
% Tree search
Dara employs an exhaustive tree search algorithm complemented by intelligent composition and structure grouping strategies to achieve robust XRD phase identification and ensure human readability. It is particularly suited for analyzing the phases in complex, multi-phase diffraction patterns, such as those obtained from powder products of solid-state reactions. Unlike the conventional use of Rietveld refinement, which primarily extracts phase structure information out of XRD patterns, Dara leverages refinement engines (e.g., BGMN) earlier in the analysis pipeline to identify candidate phases, ensuring both transparency and interpretability. Nonetheless, Dara is not intended for detailed structural refinement, such as determining atomic positions, site occupancies, or displacement parameters, which require expert knowledge and task-specific context.

The features of Dara include:
\begin{itemize}
    \item Ability to analyze complex, multi-phase diffraction patterns of solid-state reaction products, primarily focusing on phase identification
    \item Rietveld-refinement-based search algorithm that can provide good interpretability.
    \item Null hypothesis generation and testing engine to explore all possible combinations of phases. If multiple combinations fit well, they are all provided and ranked.
    \item Compatible with reference structures for multiple sources, supporting structures from both experimental and computational databases.
    \item Designed for future integration in multi-modal characterization efforts: XRD and other elemental analyses like SEM/EDS, XRF, and XPS.
\end{itemize}

%On the other hand, although Dara relies on Rietveld refinement to assess phase fitness, it is not intended for dedicated structural refinement, such as refining atomic position, site occupancy, and atomic displacement. The primary goal of Dara is phase identification: to determine which crystalline phases are present in the sample, allowing users to gain a comprehensive picture of possible phases that can fit into the diffraction pattern and be present in the sample. More detailed structural refinements, while highly valuable for extracting information such as site occupancy and atomic positions, remain downstream tasks that require task-specific, expert-driven analysis beyond the scope of this work.

\section{Methods}
\begin{figure}
    \centering
    \includegraphics[width=\linewidth]{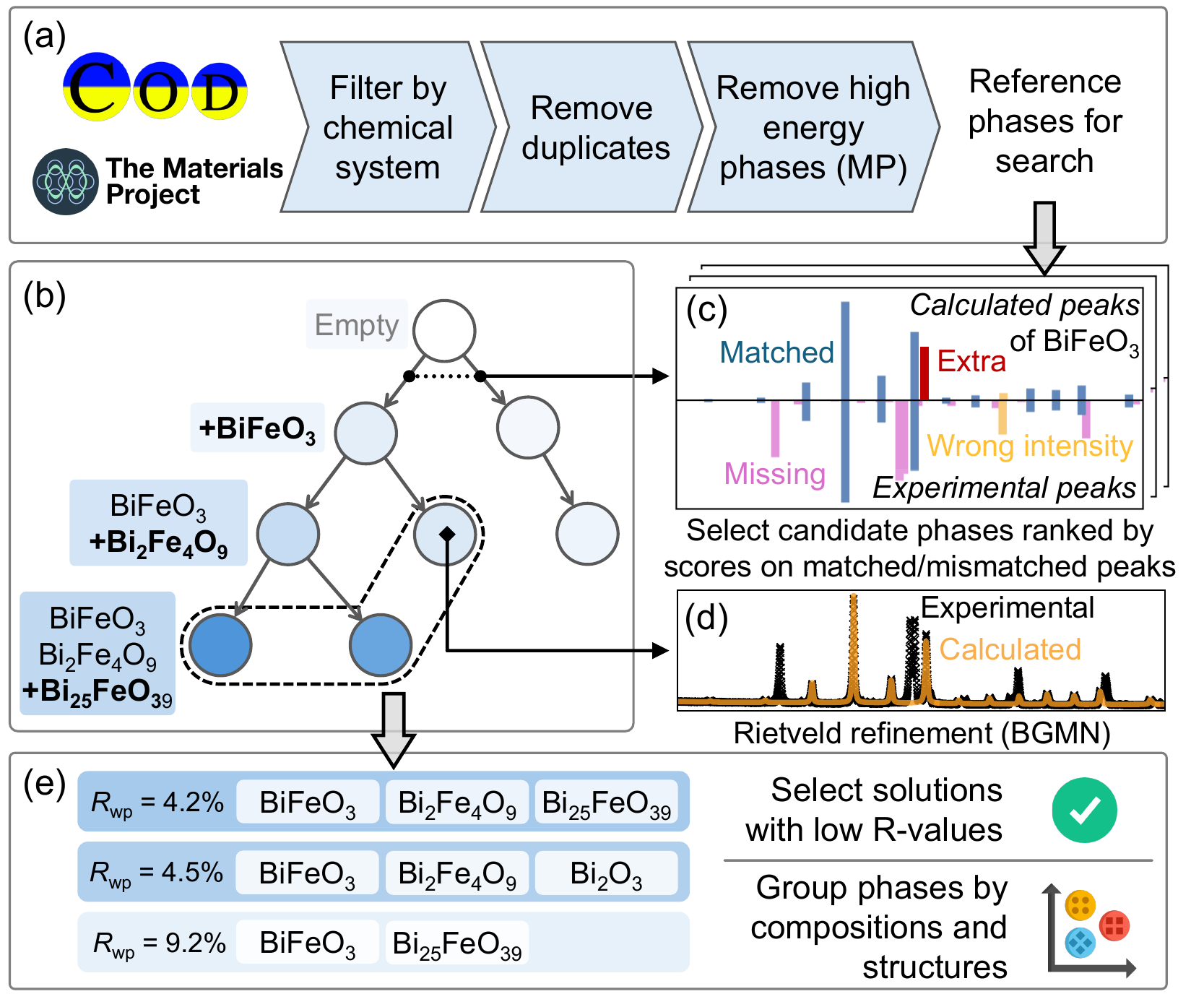}
    \caption{Overview schematic of XRD phase analysis performed with Dara. (a) The preprocessing workflow filters reference phases, which are a set of crystalline material structures, from structure databases such as the Crystallography Open Database (COD) \cite{Grazulis2012, Grazulis2009}. First, all phases within the chemical system of the input XRD pattern are selected. Then, duplicate phases are removed based on the formula and space group. High-energy phases are further filtered out using thermodynamic data from the Materials Project. The resulting phases are used as the reference phases in the downstream search routine. (b) A search tree is constructed with each node representing a phase combination, and directed edges representing the addition of one phase to the previous node’s phases. The color of each node represents the weighted profile residual ($R_{wp}$) values. Darker colors represent lower $R_{wp}$ (indicating a better fit). (c) A peak matching algorithm to quickly filter phases that can fit well to any of the remaining unmatched peaks to prune unlikely phases and save computation time. (d) Identified phases are then passed to a Rietveld refinement engine, such as BGMN. The black crosses are the experimental pattern. The orange line represents the calculated pattern output by Rietveld refinement. (e) Multiple results are extracted from the search tree and presented to the user. The results are ranked by R-values and grouped based on their compositions and structures for easier interpretation. Results with excessively high R-values are removed.}
    \label{fig:dara-components}
\end{figure}

The XRD analysis workflow of Dara is shown in Figure~\ref{fig:dara-components}, with details of each step described in the following subsections. 
In brief, the analysis begins with a set of reference phases, which are a list of crystalline material structures. These may be supplied by the user or generated automatically by Dara by filtering database entries to the sample’s element set and removing redundant entries (Figure~\ref{fig:dara-components}(a)). Dara then constructs a search tree to iteratively explore the likely reference phase combinations (Figure~\ref{fig:dara-components}(b)) and identify all phases that can be present in a sample, potentially containing multiple phases, from the reference phase set. Each node represents a phase combination, which is a subset of the reference phase set. A directed edge adds one phase to a node's phase combination, producing a child node. Because exhaustive traversal of all combinations is combinatorial and intractable, Dara employs a heuristic peak-matching score, similar to the search–match method, to prioritize promising phases and prune unlikely phases (Figure~\ref{fig:dara-components}(c)). A Rietveld refinement then evaluates the phase combination in a node to obtain accurate fit metrics (Figure~\ref{fig:dara-components}(d)). After that, Dara will expand the node by adding one more phase, creating a new node in the tree. The search ends either when a user-defined maximum number of phases in a combination is reached or when adding phases no longer improves the fit. Finally, Dara retains well-fitting phase combinations, discards poorly fitted phase combinations, and groups phase combinations by diffraction similarity and composition to present an interpretable summary of the likely phase combinations in the sample (Figure~\ref{fig:dara-components}(e)).

\subsection{Structural databases \& preprocessing}
Dara's analysis workflow starts with a set of reference phases, which are known crystalline structures, typically observed experimentally. Users can either supply the reference phases on their own, or use Dara's workflow to automatically generate and clean up the reference phases for an input XRD pattern. The workflow begins with input structure databases (e.g., COD), which undergo a series of preprocessing steps to filter redundant, problematic, and high-energy phases before being passed to the downstream tree search. Although Dara can work with reference phases without filtering, this pre-processing workflow specifically removes molecular, organic, and duplicate phases, consistent with the typical target application of analyzing inorganic crystalline solids in powder diffraction patterns.  

Duplicate phases are first identified via the structure matching algorithm in the pymatgen package\cite{ong2013python}. From each duplicate set, we select the phase characterized at a temperature closest to 20$^\circ$C and recorded earliest (i.e., with the oldest entry year) in the database. For the cases where the XRD pattern is not measured at room temperature, such as those acquired during \textit{in situ} heating, we find that the variations in lattice parameters and Debye-Waller factors are typically minor enough for Dara to accurately identify the correct phases.

To filter down the reference phases to the most plausible set, the DFT energies are retrieved from the Materials Project (MP) database \cite{jain2013commentary}. Because exact matches between experimental and DFT-optimized structures are not always available, the energy of the lowest-energy MP structure with the same space group and composition as the experimental phase is used as an approximation. Phases with an energy above the hull greater than 100 meV/atom are discarded, while phases without a corresponding MP entry are retained to avoid inadvertently excluding them.

\subsection{Tree search for phase identification}
Once the reference phase set is established, Dara initiates its core analysis by constructing a search tree. The search tree systematically explores all viable combinations of reference phases (Figure~\ref{fig:dara-components}(b)). Each node in the tree corresponds to a specific phase combination, while each directed edge represents the addition of one new phase. This process, known as node expansion, generates new phase combinations by adding one phase at a time to an existing node, thereby forming a new child node. For each node, Dara will perform Rietveld refinement to evaluate the fitness of phases to the patterns. Because the refinement is computationally expensive, Dara restricts expansion to a shorter list of promising reference phases, selected by scoring all remaining references in the chemical system with the peak-matching algorithm described later.
To avoid redundant exploration of the same phase combinations in different orders across the search tree, Dara enforces an ordering constraint during tree expansion. Each newly added phase must have a lower maximum peak intensity than that of any phase already in the node. This ensures that each unique phase combination is visited only once, progressing from the most prominent phases to the less prominent ones.

In XRD phase analysis, it is common to encounter multiple reference phases that share almost identical diffraction patterns, such as solid solutions with slight variations in composition or those with minor differences in atomic orderings. These phases almost always yield a similar fit during refinement. It is advantageous to group these phases and treat them collectively. Dara implements this strategy by using the peak-matching algorithm to quantify the similarity of diffraction patterns (described in \textit{Peak matching} section).
Furthermore, to avoid redundant refinements within a group, Dara only continues node expansion with one representative phase per group. The representative phase is chosen using the figure of merit (FoM) \cite{doebelin2015profex, lutterotti2019full}, which considers the quality factor ($1-\rho$, an R-factor variation defined in BGMN to eliminate background effect) and the lattice parameter shift ($\Delta U$). A detailed explanation of the quality factor can be found in the BGMN user guide. \cite{Bergmann2005BGMN}. The FoM is calculated as
\begin{equation}
    \begin{gathered}
        \text{FoM} = \frac{1}{(1 - \rho) + \Delta U} \\
        \Delta U = 100 \cdot \left( \frac{\vert a_\text{refined} - a_0\vert}{a_0} + \frac{\vert b_\text{refined} - b_0\vert}{b_0} + \frac{\vert c_\text{refined} - c_0\vert}{c_0}\right),
    \end{gathered}
    \label{eq:fom}
\end{equation}
where $\{a,b,c\}_\text{refined}$ represent the lattice parameters obtained from refinement, and $\{a,b,c\}_0$ represents the lattice parameters of the unrefined reference phase. The algorithm prioritizes phases that require smaller lattice parameter shifts during refinement, thereby avoiding the overfitting of solid solutions. To ensure the newly added phase improves the fit, Dara evaluates the improvement in the profile residual factors with corrected background ($R_{pb}$) \cite{Bergmann2005BGMN}, an R-factor that eliminates the background effect by subtracting the fitted background. If the improvement in $R_{pb}$ falls below a specified threshold, the node will no longer be expanded. Otherwise, the search continues until a maximum number of phases is reached. In Dara, a default $R_{pb}$ improvement threshold of 2\% is used to balance overfitting and underfitting of the pattern. The default maximum number of phases is set to 5, as peak overlap typically becomes too severe to distinguish phases beyond this point reliably. At the end of the search routine, all grouped phases, including those with lower FoM, are reported together for user consideration.

To accelerate the tree search process, Dara utilizes the Ray framework \cite{moritz2017ray} to run multiple tree node expansions and Rietveld refinements concurrently. A breadth-first search (BFS) strategy is employed, with an internal queue managing node expansion tasks. Worker processes pull tasks from the queue, with each initiating a subtree expansion that includes peak matching and Rietveld refinement. Upon completion, the expanded subtree is merged back into the main search tree managed by the master process, which then identifies and queues new nodes for further expansion. Thanks to the high scalability of Ray, Dara can run efficiently on multi-core CPUs and even across multiple nodes in high-performance computing (HPC) clusters, significantly reducing analysis latency.

\subsection{Peak matching} 
\label{subsec:peak-matching}
During node expansion, Dara performs a Rietveld refinement on each proposed phase combination. However, as the size of the reference phase set increases, the number of possible combinations, and thus refinements, would grow exponentially. To avoid unnecessary refinements, Dara uses a customized peak-matching algorithm to quickly estimate the fitness of a phase to the XRD pattern before committing to a Rietveld refinement. This algorithm finds a mapping between peaks in an experimental pattern (measured on the actual sample) and those in the calculated XRD pattern of each reference phase. Then, a heuristic fitness score is computed based on the mapping and used to evaluate the fitness of a phase to the experimental pattern. 

% The peaks in the experimental pattern are extracted using the TEIL\&EFLECH \cite{bergmann2007eflech} program within the BGMN software suite, which can accommodate patterns with significant peak overlap.

Before constructing the search tree, Dara extracts all the peaks in the experimental pattern using the TEIL\&EFLECH \cite{bergmann2007eflech} program within the BGMN software suite, which can accommodate patterns with significant peak overlap. To obtain the calculated peaks in each reference phase, Dara also performs a single-phase refinement for every reference phase. The goal of this step is to generate calculated diffraction peaks for each phase, which serve as the basis for comparison with the experimental pattern during peak matching. Peaks in both experimental and calculated patterns are classified into four categories (Figure~\ref{fig:dara-components}(c)): matched, wrong intensity, missing, and extra. The matched peaks refer to peaks that appear in both the calculated and experimental patterns at nearly the same position and with similar intensities. The wrong intensity peaks refer to those that appear in similar positions but with very different intensities (i.e., by a factor greater than five, which is the default used in Dara). The missing peaks refer to those that appear only in the experimental pattern but not in the calculated pattern, which can indicate either a poor fit or the existence of other phases. The extra peaks refer to those that appear only in the calculated pattern but not in the experimental pattern. This sometimes occurs when the Rietveld refinement procedure determines that it is a mathematically more optimal fit to match only the major peaks in an experimental pattern while leaving some minor peaks unmatched. Extra peaks in the calculated pattern typically indicate a structural discrepancy, such as symmetry breaking from different atom ordering, between the actual material and the reference phase, making the latter less favorable for selection in phase identification. After classification, a heuristic score for each reference phase is calculated based on the intensity of peaks in different categories, as
\begin{equation}
    \mathrm{Score} = \frac{\sum I_\mathrm{matched} + \sum I_\text{wrong intensity} - 0.1 \sum I_\mathrm{missing} - 0.5 \sum I_\mathrm{extra}}{\sum I_\mathrm{exp}}
    \label{eq:score},
\end{equation}
where $I$ is the intensity of each peak, with $I_\mathrm{matched}$, $ I_\text{wrong intensity}$, and $I_\mathrm{missing}$ referring to the intensity of matched, wrong-intensity, and missing peaks in the experimental peak list, respectively. $I_\mathrm{extra, calculated}$ refers to the extra peaks in the calculated peak list. The score is normalized by $I_\mathrm{exp}$, the total intensity of all experimental peaks. The score function is designed such that the presence of missing and extra peaks penalizes (i.e., decreases) the score. In contrast, a greater number of matched and wrong-intensity peaks increases the score. The coefficients were determined through a heuristic process and can be adjusted. 

By calculating scores for all reference phases, Dara can quickly identify phases that potentially have a good fit and warrant further Rietveld refinement, which is a slower but more accurate process. When a node is expanded during the tree search, the missing experimental peaks are extracted by peak matching algorithm, indicating none of the phases in the node can account for these peaks. Afterwards, a new peak matching is performed for each reference phases' calculated peaks against the missing experimental peaks. The score is then calculated from the peak matching result to measure the fitness of a reference phase. In practice, only a small subset of phases achieves high scores, indicating good fitting. The majority show poor agreement due to the significant difference between the calculated pattern and the experimental XRD pattern. Instead of applying a fixed threshold, Dara dynamically determines a threshold by detecting the transition between high- and low-scoring phases. This is achieved by analyzing the cumulative percentile distribution of scores. The maximum of its second derivative (the inflection point) marks where the scores shift most sharply from good to poor, which can be seen as a boundary between good and poor fitting. Only phases with a score higher than this threshold will be added to the phase combination to create new nodes, where Rietveld refinement is performed to evaluate the fitness more accurately.

In addition to quickly estimating the fitness of a phase, Dara also employs the peak matching algorithm to group phases with similar XRD patterns during node expansion. To this goal, Dara runs the aforementioned peak-matching algorithm between two calculated patterns from two different reference phases and classifies each peak into one of the four categories. Then, a Jaccard index is used to calculate the similarity between the two computed patterns, as
\begin{equation}
    \text{Similarity(Pattern 1, Pattern 2)}=\frac{\sum I_\text{matched+wrong intensity}^\text{Pattern 1}+\sum I_\text{matched+wrong intensity}^\text{Pattern 2}}{\sum I^\text{Pattern 1}+\sum I^\text{Pattern 2}}.
\end{equation}
For each node expansion, the pairwise similarity between all newly added phases is calculated, forming a similarity matrix. Then, an agglomerative clustering algorithm \cite{ziegel2003elements} is applied to the similarity matrix to group phases that exhibit nearly identical calculated patterns. We use the clustering algorithm implemented in scikit-learn \cite{scikit-learn}, with a default similarity threshold of 0.9. Only one representative node in each node group is selected with FoM and continues the node epxansion. Others will be considered as alternative structure solution, as described in the \textit{Tree search for phase identification} section.

\subsection{BGMN Rietveld refinement engine}
Rietveld refinement in Dara is performed with the BGMN \cite{bergmann1998bgmn, taut1998seifert} package, as shown in Figure\ ~\ref{fig:dara-components}(d). The BGMN binaries (v4.2.23) are used as compiled and supplied by the Profex team. \cite{doebelin2015profex}. Two sets of refinement parameters are used in Dara: (1) search refinement parameters, which are used during the phase search stage and restrict peak broadening and preferred orientation to avoid overfitting, and (2) final refinement parameters, which allow a wider range of adjustments (e.g., larger peak broadening and preferred orientation). 

For the refinement parameters in the phase search stage, a maximum of 1\% lattice strain is allowed on each phase during refinement. A peak model with Cauchy square broadening ($r_p$=4) is used to describe the peak shape. The width parameter, $k_1$, is constrained between 0 and 0.01. The second width parameter, $k_2$, is fixed to 0. The crystalline size parameter, $B_1$, is constrained between 0 and 0.005. The weight fraction of each phase is calculated directly by normalizing the scale factor (GEWICHT) of all the phases in each pattern. The sample displacement factor, EPS2, is constrained between -0.05 and 0.05. 

After the phase search stage, a finer refinement step is conducted on each searched phase combination to obtain a more accurate fit of the XRD peaks and determine phase fractions. In this refinement, up to 1\% lattice strain is allowed for each phase. Peak shapes are modeled using a Cauchy-squared broadening function with a profile parameter \( r_p = 4 \). The peak width parameter \( k_1 \) is constrained between 0 and 0.01, while \( k_2 \) is fixed at 0. The crystallite size parameter \( B_1 \) is constrained between 0 and 0.05. Preferred orientation is accounted for using a fourth-order spherical harmonic function (SPHAR4). Phase weight fractions are calculated by normalizing the scaling factor (GEWICHT) across all identified phases in a given pattern. The sample displacement parameter EPS2 is constrained between –0.05 and 0.05 to correct for sample height effects.

\subsection{Result representation and compositional grouping}
At the end of the tree search, Dara collects all leaf nodes. These phase combinations can span a wide range of fit quality, from poor to good, so further filtering is required. To retain only meaningful results, Dara applies the Jenks natural break detection algorithm \cite{jenks1967data}, which clusters phase combinations with similar quality factor ($1-\rho$) together by minimizing variance of quality factors within each cluster. Dara then reports the cluster of solutions with the lowest quality factor as the final result of phase combinations that can fit the pattern while discarding those with a clearly poor fit.

For easier human interpretation, the identified phases are also further grouped by composition. This compositional grouping is performed using an agglomerative clustering algorithm \cite{ziegel2003elements}. For each composition group, the representative composition is chosen as the integer composition closest to the average composition. If no integer composition is available, the selected composition is the one closest to the average composition of the group. By clustering compositionally similar phases, this approach reduces redundancy in the reported results and highlights the most relevant, representative phases.

Additionally, thanks to the peak-matching algorithm, the search results include information describing any unmatched peaks. If a given phase combination cannot account for all the peaks in the experimental pattern or introduces extra peaks, the peak-matching algorithm flags these discrepancies. This provides an additional measure of the quality of phase identification, allowing users to assess the extent to which the proposed phase combination accurately explains the experimental data. The reported differences serve as indicators of potential missing or misidentified phases, guiding further refinement in phase selection and structure analysis.

\section{Results}
To evaluate Dara's performance in analyzing real-world XRD patterns, we create several benchmark scenarios of real powder diffraction patterns. The results of these tests are described in the following sections.

\subsection{Benchmarking on a dataset of commercial precursor mixtures}
For the first test case, we construct a benchmark XRD pattern dataset by mixing commercial precursor materials (oxides and carbonates) in varying ratios. Ten crystalline, single-phase precursors are selected and used to prepare 10 binary (two-component) and 10 ternary (three-component) mixtures. In each mixture, the precursors are randomly selected to constitute between 10 wt\% and 90 wt\% of the total mass, yielding a wide range of peak intensities to assess Dara’s performance across different weight fractions, as illustrated in Figure~\ref{fig:precursor_mixture_benchmark}(a). Detailed preparation protocols are provided in Supplementary Note~\ref{si:preparation_mixture}. Each sample is measured using two scan settings: a short scan (2 minutes, low quality) and a longer scan (8 minutes, medium quality) between 10\textdegree~and 100\textdegree, enabling evaluation across different noise levels.

\begin{figure}[H]
    \centering
    \includegraphics[width=\linewidth]{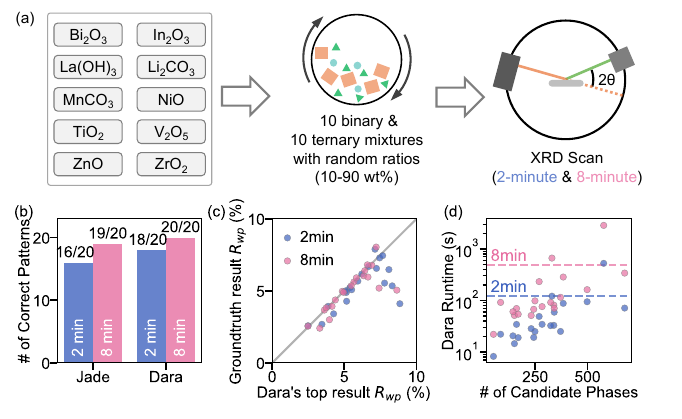}
\caption{Preparation of the precursor mixture dataset and benchmarking results.
(a) Schematic illustration of the procedure for generating the precursor mixture dataset. Ten commercial precursors are randomly selected and mixed at varying ratios between 10 wt\% and 90 wt\%, resulting in 10 binary and 10 ternary precursor mixtures. XRD patterns are collected using a benchtop diffractometer under two scanning programs (2 minutes and 8 minutes) to produce datasets of different measurement qualities.
(b) Comparison of correctly indexed patterns by Jade and Dara. Correct means the analysis method successfully identifies all the precursor phases without any spurious phases. Blue bars indicate patterns scanned using the 2-minute (low-quality) program, while pink bars represent the 8-minute (medium-quality) scans. The top of each bar shows the number of correct predictions compared to the total number of patterns for that scan type.
(c) Relationship between the $R_{wp}$ values from Rietveld refinement using groundtruth phases and Dara’s top result (represents the solution of lowest $R_{wp}$ that Dara can find). The ground-truth phases are the precursor phases added during sample preparation. Blue and pink dots correspond to 2-minute and 8-minute scans, respectively.
(d) Dara's runtime per pattern as a function of the number of reference phases in the database. Blue and pink dots represent 2-minute and 8-minute scans, respectively. Dashed horizontal lines mark 2 and 8 minutes on the time axis, corresponding to the measurement time to obtain these patterns in the diffractometer. The time is measured on a workstation with Intel(R) Core(TM) i9-10920X CPU @ 3.50GHz.}
\label{fig:precursor_mixture_benchmark}
\end{figure}

We compare Dara’s phase identification performance against Jade\cite{jade2019}, a widely used commercial software for powder XRD analysis that integrates both phase identification and basic full-profile fitting. Parameters for both methods are described in Supplementary Note~\ref{si:benchmark_params}. We count the number of correct patterns identified by each analysis method, where a pattern is considered “correct” if the result includes all precursor phases added when preparing the sample and excludes any spurious phases. One exception is made for the pattern composed of 30 wt\% NiO, 30 wt\% \ce{Bi2O3}, and 40 wt\% \ce{Li2CO3}. In this case, both Jade and Dara detect a minor \ce{Bi2O2(CO3)} phase, present at approximately 3 wt\%, which may result from a reaction between \ce{Bi2O3} and \ce{Li2CO3} or \ce{CO2} in the air during mixing. Although this phase exhibits distinct peaks in the XRD pattern, we choose to exclude it from our evaluation due to the lack of supporting evidence beyond the diffraction data. Results are shown in Figure~\ref{fig:precursor_mixture_benchmark}(b). For the 2-minute scans, Jade misclassifies 4 out of 20 patterns: either missing a phase or introducing spurious ones. Dara misclassifies only two patterns, both of which result from missing a phase. For the higher-quality 8-minute scans, both methods show improved accuracy due to better signal-to-noise ratios. In this case, Dara still outperforms Jade, identifying all 20 patterns successfully, while Jade still fails in 2 cases. Detailed analysis of Dara's and Jade's phase identification results for all samples is provided in Supplementary Note~\ref{si:result_mixture}.

To further understand the fitness of the phases selected by Dara, we plot the weighted profile residuals ($R_{wp}$) of Dara's top result (the result with the lowest $R_{wp}$) against the $R_{wp}$ produced from human-performed Rietveld refinement. For each precursor used in constructing the dataset, the groundtruth reference phase is carefully hand-picked by humans from the structural database and a very good fit is obtained with the XRD pattern of the single precursor, as shown in Supplementary Note \ref{si-note:single-phase-refinement}. Hence, these reference phases are considered sufficiently similar that they can be used as an approximation to the actual phase in the sample (groundtruth). The ICSD IDs of these phases are listed in Supplementary Table \ref{si-table:icsd-number}. Figure~\ref {fig:precursor_mixture_benchmark}(c) compares the $R_{wp}$ of refining with the known groundtruth phases versus Dara's top-ranked refinement outcome, as listed in Supplementary Table~\ref{si-tab:precursor_rwp}. Across the dataset, Dara consistently returns solutions with $R_{wp} < 10\%$, reflecting good-quality fits and generally aligns with the groundtruth results. However, in some cases, the groundtruth $R_{wp}$ is lower than Dara’s. This discrepancy can arise from the two-stage nature of our approach and the varied refinement parameters used at different stages, as described in the \textit{Method} section. After phase searching, Dara performs the final refinement using only the representative phases with the highest Figure of Merit (FoM). In some cases, the phases that fit best during the constraint search stage are not the ones that would fit best for the final refinement, where Rietveld refinement is allowed to adjust more parameters to achieve a better fit. Even so, the selected phases can still be regarded as valid matches to the experimental pattern, as the differences primarily arise from variations in peak intensity rather than the appearance or disappearance of specific peaks.
These cases underscore the ambiguity in XRD analysis: multiple, subtly different phases can produce comparably good fits under different refinement parameters, and distinguishing between them often requires careful inspection and, in some cases, additional characterization.

Finally, we evaluate Dara’s use of computational resources. Different from Jade’s fast, heuristic search-match algorithm, which completes in seconds, Dara relies on a more exhaustive and detailed evaluation of phases that involve hundreds of Rietveld refinements for a single pattern. This approach admittedly requires more computational time but is better suited to handling complex, multi-phase patterns. Figure~\ref{fig:precursor_mixture_benchmark}(d) shows that runtime scales with the number of reference phases. Dara processes 2-minute scans (blue dots) slightly faster than 8-minute scans (pink dots), mainly due to the fewer 2-theta angular steps measured in a shorter scan, which reduces the refinement time. This is because the refinement requires the computation of intensity at every angular point in the pattern to calculate the error. Several runtime optimization strategies are implemented in Dara: (1) an integrated heuristic search-match step to filter unlikely phases before committing to a full-profile Rietveld refinement that can be time-consuming; (2) grouping of structurally similar phases to avoid redundant refinements on XRD-equivalent variants (e.g., doped or vacancy-modified forms); and (3) parallel execution using the Ray framework \cite{moritz2017ray}, enabling scaled-up deployment on multi-node computing clusters. As a result, the typical runtime per pattern remains shorter than the actual measurement time (2 or 8 minutes), marked as dashed horizontal lines in Figure~\ref{fig:precursor_mixture_benchmark}(d).

\subsection{Benchmarking on pairwise reaction product dataset}
\begin{figure}
    \centering
    \includegraphics[width=1\linewidth]{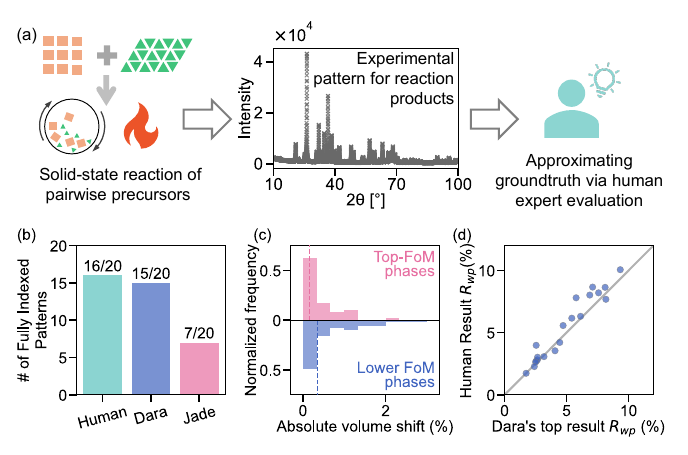}
    \caption{Preparation of the solid-state reaction dataset and benchmarking results. (a) Schematic illustration of the workflow for preparing the reaction dataset and approximating ground truth solutions through human expert evaluation. (b) Number of XRD patterns that have all peaks indexed in the analyses conducted by human experts, Dara, and Jade. (c) Comparison of the absolute lattice volume shifts after Rietveld refinement between the phases with the highest FoM scores (pink bars) and other phases with lower scores (blue bars) within each phase group found by Dara. A phase group refers to a set of phases that yield similar XRD patterns and are therefore expected to provide a comparable fit to the experimental peaks. The x-axis represents the absolute value of lattice volume shift (in \%), and the y-axis shows the normalized frequency (with the sum of frequency set to 1). Pink bars on top indicate the volume shifts of the most appropriate phases selected by Dara in each phase group, based on a figure of merit (FoM) that incorporates both the quality factor $(1 - \rho)$ and lattice shift $(\Delta U)$. Blue bars represent all other phases with lower FoM, which Dara did not choose to continue the search but considers as alternative phases to the FoM-selected phases in the result. The median for each histogram is shown as a dashed line in the plot. (d) Correlation between the $R_{wp}$ values from human-expert Rietveld refinement and Dara’s best-fit solution (i.e., the one yielding the lowest $R_{wp}$ identified by Dara).}
    \label{fig:benchmark-binary}
\end{figure}

The second benchmark is designed to evaluate Dara’s performance on a dataset composed of the products of inorganic solid-state reactions. In these reactions, both precursors and products are typically solid powders. Due to the slow kinetics of solid-state diffusion, products often consist of off-stoichiometric solid solutions, unreacted precursors, and metastable intermediate phases. \cite{mcdermott2023assessing} These complex multiphase mixtures, frequently exhibiting significant variations in composition and structure, pose a major challenge for XRD analysis, despite XRD being a key technique for characterizing the outcomes of solid-state reactions. To this end, we construct a dataset comprising 20 samples from reactions between pairs of precursors chosen from 21 commonly used precursors, including oxides, carbonates, phosphates, and oxalates. Reaction temperatures are selected based on Tammann’s rule \cite{merkle2005tammann}, which estimates the onset temperature of solid-state reactions to be roughly two-thirds of the lowest melting point among the precursors. Since this is often a somewhat low temperature, many reactions are incomplete with partially unreacted precursors. Such a multi-phase mixture tests Dara's ability to make correct phase assignments for realistic samples that are often encountered in the exploratory phases of synthesis. The procedure to obtain the dataset is summarized in Figure~\ref{fig:benchmark-binary}(a). All experiments are performed in A-Lab, a fully automated synthesis platform equipped with robotic arms.\cite{szymanski2023autonomous} Additional details on sample synthesis and XRD characterization are provided in Supplementary Note~\ref{si:preparation_binary}. The resulting XRD patterns are analyzed by a human expert using a typical XRD analysis and refinement approach, incorporating the PDF-5+ database and suggestions from Dara, to arrive at chemically reasonable interpretations (see Supplementary Note~\ref{si:human-eval} for details). For benchmarking the automated tools, both Jade and Dara are also run in a fully unsupervised mode, without human intervention.

We first examine the accuracy of the phases identified by Dara. In solid-state reactions, it is common to form phases with compositions and lattice parameters that differ from those of the reference phases in the structure database. Sometimes, phases with a new structure that has not been included in the structure database can be encountered. In such cases, a good XRD pattern analyzer should still function effectively and identify all plausible known phases that can provide a good fit to the pattern. To evaluate Dara’s ability to handle such off-standard phases, we compare the phases identified by Dara, Jade, and the human expert across 20 XRD patterns with results shown in Figure \ref{fig:benchmark-binary}(b). For each method, we count the number of patterns that have all peaks indexed. Only the top solution returned by Dara is considered. The human expert produces the most fully indexed patterns (16 out of 20), followed closely by Dara (15 out of 20), with four cases failing in the same manner as in the human's analysis, potentially due to some unknown phases absent from the structure databases. The additional failure of Dara (2\ce{Fe3O4} + 3\ce{Y2O3} @ 1000 \textsuperscript{\textdegree}C) occurs because it identifies the wrong polymorph of \ce{Y2O3} (\hmn{Fm-3m} instead of the ground-state \hmn{Ia-3}), leading to a clearly unmatched peaks at around 20$^\circ$. This is because the major phase, \ce{YFeO3}, also has a minor peak at 20$^\circ$, causing Dara's peak matching algorithm to mistakenly treat the 20$^\circ$ peak as ``matched'' and deprioritize searching for phases that have a 20$^\circ$ peak. Jade, on the other hand, only manages to index 7 of the 20 patterns fully. We attribute this mainly to the fact that many phases formed in solid-state reactions differ from the reference phases in the structure database, which may make them difficult to capture via the search-match method. However, these phases can be captured by the full-profile fitting process, which models various XRD-related effects (e.g., sample displacement and broadened peaks) during the fitting phases. As for the precursor mixture dataset, we analyze the runtime of Dara for each pattern relative to the number of reference phases (Supplementary Figure~\ref{si-figure:binary_reaction_runtime}). Since most samples in this benchmark contain fewer elements, there are fewer reference phases, and runtimes are generally lower than in the precursor mixture dataset. Most analysis workflows complete in under two minutes. Four samples required 2-8 minutes, while two samples require more than 8 minutes, out of the total twenty patterns.

Dara can identify multiple phases that yield comparably good fits to a diffraction pattern, helping users recognize the range of possible interpretations before drawing conclusions from XRD data. For example, in the reaction \ce{Cr2O3} + 2 \ce{MnO} at 1100 \textsuperscript{\textdegree}C, two spinel phases with distinct compositions are detected: \ce{MnCr2O4} and \ce{CrMn_{1.5}O4}. Due to the similar structure factor of Mn and Cr and the close lattice parameters ($a=8.435\text{\AA}$ and $8.479\text{\AA}$, respectively), the XRD patterns of these two phases are hard to distinguish through a several-minute XRD scan on a lab X-ray diffractometer. For this example, Dara’s refinement indicates a spinel phase with a lattice parameter of $a=8.454\text{\AA}$, suggesting a composition intermediate between the two spinels, potentially a Mn:Cr ratio of 1:1, which is indeed the Mn:Cr ratio in the precursor of this sample.
This ability to identify potential fits is particularly useful for analyzing the reaction products when attempting to synthesize materials predicted by computational screening. By including the computed structure in the reference phase set, Dara is able not only to tell if the computed structure has a good match to the pattern but also to list all the known structures that have a good match in the structure database. If both the computed structure and the known phases can fit equally well to the pattern, Dara will catch the user's attention and suggest additional analysis (e.g., SEM/EDS) to verify elemental compositions.

In Dara’s search, phases are grouped based on the similarity of their diffraction patterns. Phases within the same group are treated as effectively indistinguishable by XRD and can produce a similar fit to the given experimental pattern. From each group, a representative phase is selected using a figure of merit (FoM), adapted from Lutterotti et al. \cite{lutterotti2019full}. This FoM, defined in Equation~\ref {eq:fom} (see Methods), combines a fitness term ($1 - \rho$), which evaluates the overall fit quality, and a lattice shift term ($\Delta U$), which measures the change in lattice parameters $a$, $b$, and $c$ before and after the refinement. %By incorporating both the fit quality and the extent of lattice parameter shift, the FoM captures not only how well a reference phase matches the diffraction pattern, but also the structural deviation between the reference and refined phases, which can be an indicator of potential compositional differences between the reference phase and the actual phase present in the sample due to the formation of solid solution.
To demonstrate the effectiveness of this selection criterion, we computed the lattice volume shifts during refinement for both top-FoM and lower-FoM phases in each phase group (Figure~\ref{fig:benchmark-binary}(c)). The top-FoM phases (pink bars) exhibit smaller shifts, with a median of 0.15\%, meaning the lattice volume is adjusted by only 0.15\% during Rietveld refinement to align with the experimental peak positions. In contrast, the lower-FoM phases (blue bars) show larger shifts, with a median of 0.35\%. This difference highlights that the FoM score reflects not only the fit quality (measured by quality factor $1-\rho$) but also the changes in the lattice parameters, which often indicates compositional variation between the actual and reference phases, which is a common product in the reaction products of the solid-state reactions, primarily due to the formation of off-stoichiometric phases. This FoM prioritizes reference phases with lattice parameters more closely matching those of the sample, which can aid in estimating the composition of solid solutions present.
A similar analysis of the precursor mixture dataset (Supplementary Figure~\ref{si-figure:lattice_shift_precursor}) reveals the same trend. The top-FoM phases have a median lattice volume shift of 0.14\%, while the lower-FoM phases show a median of 0.40\%. Compared with the pairwise reaction dataset, the top-FoM phases in the precursor mixture dataset display slightly smaller shifts, whereas the lower-FoM phases deviate more. This is because the precursors used in the precursor mixture dataset have often been thoroughly characterized in the structure database and reported with larger ranges in lattice parameters, primarily due to the different synthesis methods and measurement conditions. Typically, at least one reference phase closely matches the lattice parameters of the sample, resulting in a slight lattice shift for the top-FoM phase. Other reported phases may require larger lattice adjustments to align with the experimental peaks, yet still achieve a comparable fit owing to their nearly identical crystal structures. As a result, Dara regards these phases as equally valid matches to the diffraction pattern. Distinguishing them requires more dedicated analysis of the XRD pattern or additional characterization.

%This difference can be attributed to the density of reported entries for each material in the structural database: well-characterized precursors, which are the phase in the precursor mixture samples, often have many reported phases that differ only by minor structural variations and lattice parameters, resulting in a wider spread of lattice parameters and producing similar fits to the pattern. This increased popularity results in the likelihood for FoM to find the phases that can result in good fitness while maintaining a small lattice volume shift. However, as more phases can match the pattern due to the similarity of XRD patterns produced by these reference phases, the FoM of other non-selected phases will spread more broadly. This underscores the necessity to consider lattice parameters during selecting the most possible phase, as many phases can produce similar fit and lattice shift can be a good descriptor to measure differences between the pattern and reference phase.
%These redundant entries increase the likelihood of mismatched references, inflating the lattice volume shifts in lower-FoM phases, while FoM selection consistently favors the best-matching entry.

Finally, we compare refinement outputs from Dara and human experts. Figure~\ref{fig:benchmark-binary}(d) shows a scatter plot of final $R_{wp}$ values: Dara’s top (x-axis) versus the expert’s (y-axis). The points generally follow the $y = x$ line, indicating comparable fitting quality. Interestingly, human-derived $R_{wp}$ values are often slightly higher, particularly in higher-$R_{wp}$ cases. This may result from the different refinement strategies: human experts emphasize physical interpretability and carefully adjust refinement parameters step by step; Dara employs a consistent, general-purpose refinement setup to achieve a good fit while minimizing the risk of overfitting. Although Dara’s Rietveld refinements are not intended for detailed microstructural analysis (e.g., grain size, strain, or site occupancy), they can provide reliable phase identification (indicated by the peak fitness) and estimation of phase fractions. Additionally, they can serve as excellent starting points for downstream, more detailed analysis to obtain information such as grain size, strain, and atomic occupancy from the XRD patterns.

\subsection{Characterizing the ambiguity of XRD patterns} \label{sec:sw}
In practice, it is often hard to map the XRD pattern deterministically to the crystal structures due to the presence of multiple phases, instrument resolution limitations, peak overlap, and measurement noise. \cite{schreiner1982systematic} Multiple phases, with different compositions or slight variations in atomic arrangement, can produce almost indistinguishable XRD patterns. The limited resolution of laboratory X-ray diffractometers further amplifies this challenge. For example, diffraction peaks often overlap with one another or become obscured by background noise, making it challenging to identify individual phases confidently. In such cases, XRD alone will not be adequate to disambiguate these possibilities. Despite these limitations, lab-based XRD remains one of the most widely used tools for characterizing inorganic crystal samples. 

We illustrate this challenge using a sample synthesized in our lab, shown in Figure~\ref{fig:sw}. More details about sample preparation are described in Supplementary Note~\ref{si:sw}. It is the product of a solid-state reaction involving six elements: Li, Na, Al, Si, Co, and O. Due to the use of a Cu anode X-ray source and the presence of Co in the sample, significant background arises from secondary fluorescence, increasing the noise level. This results in a lower signal-to-noise ratio, making the XRD pattern more ambiguous for analysis. Nonetheless, although non-ideal for analysis, such patterns are typical of those encountered when analyzing real-world samples.

As shown in Figure~\ref{fig:sw}, Dara identifies four groups of phase combinations (solutions) for this sample, each consisting of three phases. The $R_{wp}$ for all four solutions ranges closely between 2.20\% and 2.33\%. Two major phases are consistently present across all four solutions. The first is a nepheline-type phase with a composition close to \ce{NaAlSiO4}, for which Dara finds 11 matching reference phases in the structure database. The second is a series of solid solutions spanning the \ce{LiCoO2}–\ce{LiAlO2} tie line. These phases share similar lattice parameters (\ce{LiAlO2}, \hmn{R-3m}: $a$ = 2.800\AA, $c$ = 14.216\AA; \ce{LiCoO2}, \hmn{R-3m}: $a$ = 2.816\AA, $c$ = 14.054\AA) and can form across a wide range of compositions \cite{khan2006synthesis,buta1999phase}. Given this, it is difficult to determine the exact composition solely from the XRD measurement.
While the two major phases remain the same across the four solutions, the third, minor phase picked by Dara varies and contains structurally and compositionally distinct phases: \ce{SiO2} (\hmn{P3_221}), \ce{Co11O16}/\ce{Co2SiO4} (\hmn{Fd-3m}) (7 matched phases), \ce{Al2CoO4} (\hmn{Fd-3m}) (25 matched phases), and \ce{NaCo3O4}. Since they have dissimilar diffraction patterns, these phases are grouped into four solution groups. For example, despite compositional differences between \ce{Co11O16} and \ce{Co2SiO4}, they are grouped due to their similar XRD patterns. As illustrated in the bottom row of Figure~\ref{fig:sw}, each phase has its distinct diffraction peaks, but all exhibit a prominent peak that can be matched to a peak in the pattern at around 36.5\textsuperscript{\textdegree}, as flagged with an orange triangle in Figure~\ref{fig:sw}. However, due to substantial peak overlap between these minor phases and the major phases, it is challenging to determine which one is the actual phase that contributes to that peak. Detailed phase lists for each solution and composition groupings are provided in Supplementary Note~\ref{si:sw_result}.

\begin{figure}
    \centering
    \includegraphics[width=0.7\linewidth]{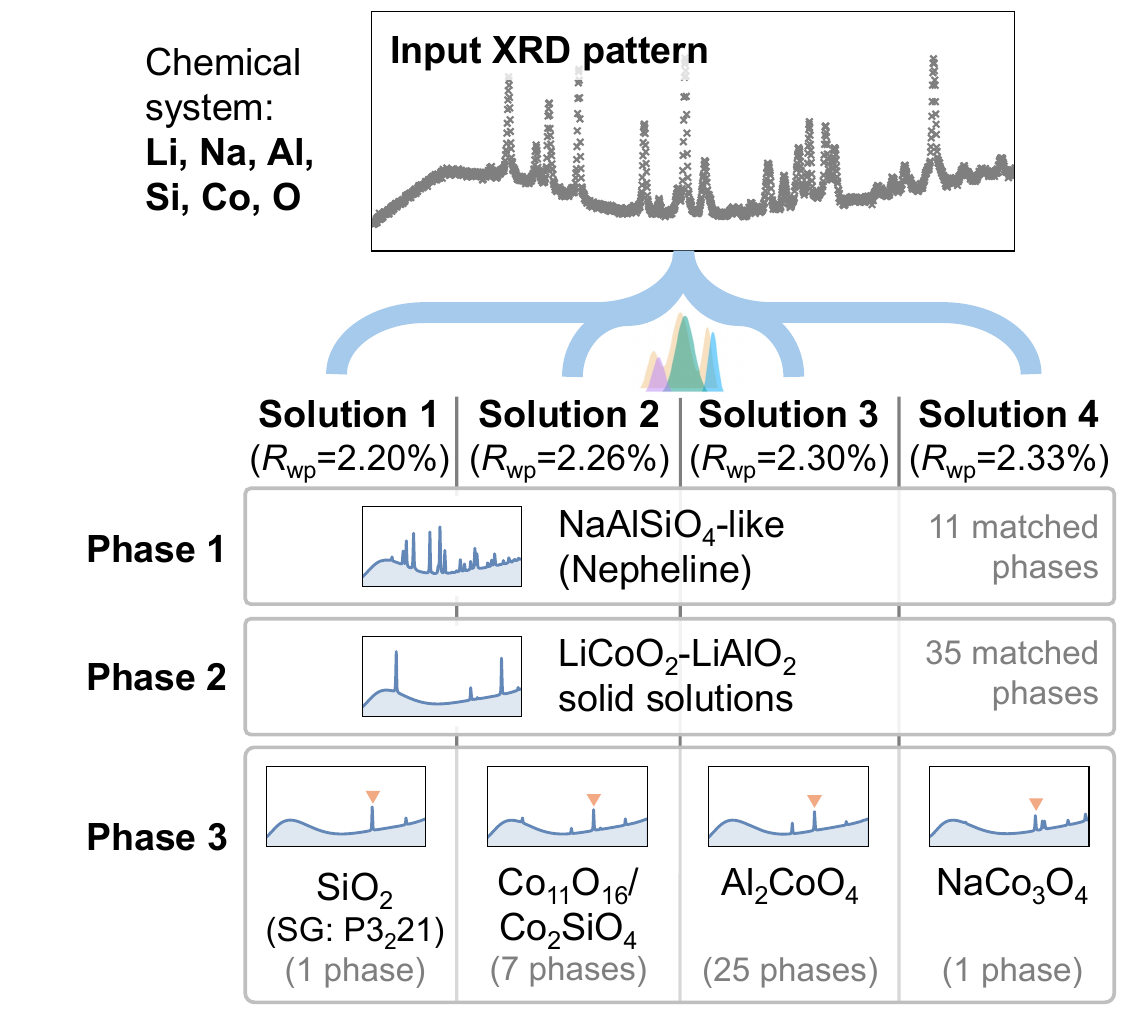}
    \caption{Example of multiple phase solutions identified by Dara for an experimental solid-state reaction sample. The raw XRD pattern and its corresponding chemical system are supplied to Dara. After searching, four solutions are found to fit the pattern similarly well, all of which contain three phases. The calculated patterns for each phase are displayed in the corresponding boxes in the plot. Phases 1 and 2 are shared across all the solutions, which are groups of \ce{NaAlSiO4} with Nepheline structure (11 phases) and \ce{LiCoO2}-\ce{LiAlO2} solid solutions (35 phases), respectively. Phase 3, however, includes four possible phases that differ greatly in structure/composition: \ce{SiO2} (ICSD \#155249), \ce{Co11O16}/\ce{Co2SiO4} structure family (7 phases), \ce{Al2CoO4} structure family (25 phases), and \ce{NaCo3O4} (ICSD \#163993), indicating that further compositional characterization may be necessary. The common peak at around 36.5\textsuperscript{\textdegree} is marked with an orange triangle in the plot.}
    \label{fig:sw}
\end{figure}

% In addition to grouping phases based on their diffraction patterns, we also organize all phases within a solution by composition to enhance readability and clarity. Each composition group is assigned pa representative composition. Detailed phase lists for each solution and composition groupings are provided in Supplementary Note~\ref{si:sw_result}.

\section{Discussion}
\begin{figure}
    \centering
    \includegraphics[width=0.9\linewidth]{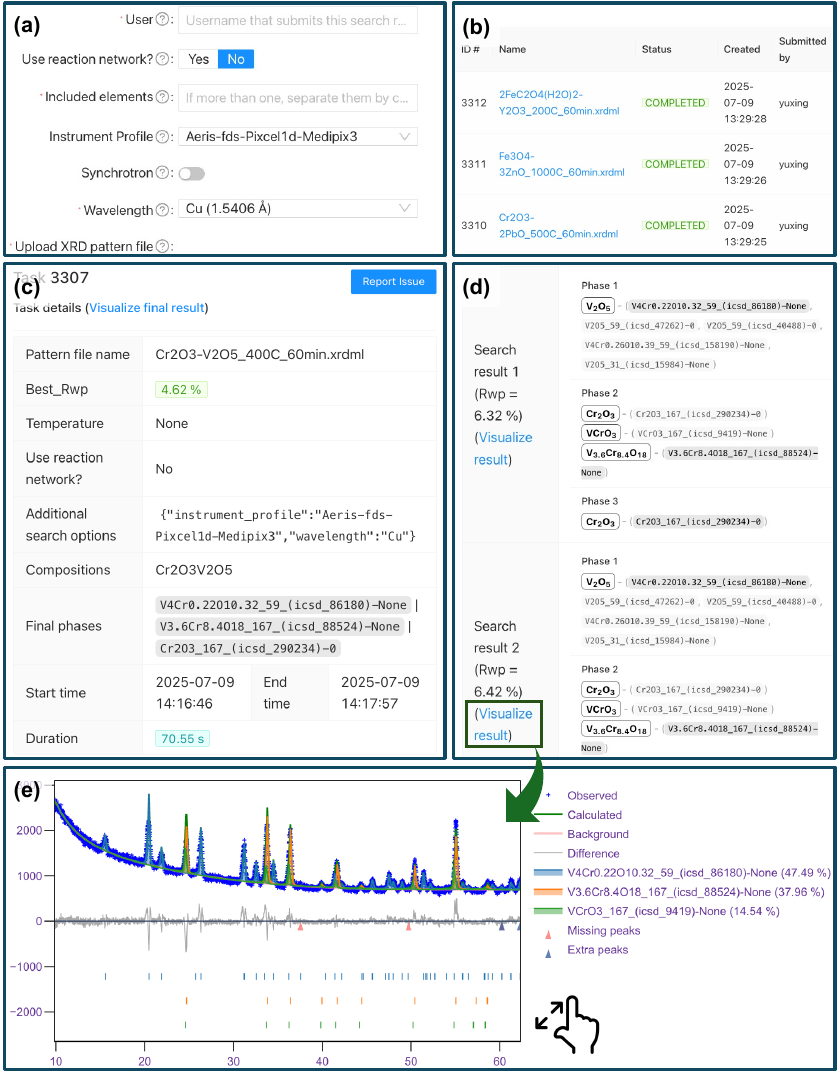}
    \caption{Screenshots of the Dara web interface. (a) Analysis job submission page, where users input the pattern, elements that can exist in the pattern, and diffractometer information. (b) Overview page for viewing the status of and accessing each analysis job. (c) Result detail page with a summary of the job's outcome, including analysis parameters, the best result's $R_{wp}$, and the most probable phases. (d) Result detail page with all solutions and phases found by Dara. (e) Interactive plot to visualize the refinement produced by Dara.}
    \label{fig:dara_service_web}
\end{figure}

Dara has been deployed in both the autonomous laboratory (A-Lab) \cite{szymanski2023autonomous} and standard research lab environments at Lawrence Berkeley National Laboratory. Through an internal web interface and API, Dara offers a user-friendly interface for automatically analyzing the XRD data. As of July 13, 2025, it has processed 2,453 unique XRD patterns from our research group and external collaborators. Screenshots of the internal application are shown in Figure~\ref{fig:dara_service_web}, including submission, overview, result details, and refinement plot visualization. This browser-based application makes Dara accessible to users who are not familiar with programming and XRD analysis.

\begin{figure}
    \centering
    \includegraphics[width=0.6\linewidth]{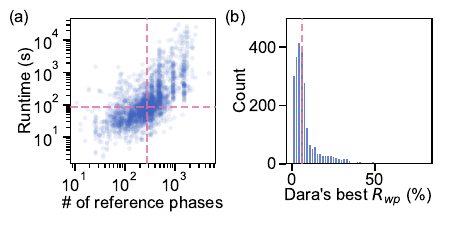}
    \caption{Statistical analysis of patterns running on our internal Dara web-based application at Lawrence Berkeley National Laboratory. (a) The relationship between the runtime of analyzing a pattern and the number of reference phases. The vertical and horizontal dashed pink lines represent the median value for runtime and number of reference phases, respectively. The runtime is measured on a workstation with Intel(R) Core(TM) i9-10920X CPU @ 3.50GHz. (b) Histogram of the best $R_{wp}$ values of each pattern obtained at the end of Dara search. The pink vertical line represents the median $R_{wp}$.}
    \label{fig:dara_service}
\end{figure}

Statistical analysis was performed on all Dara searches conducted through this web-based platform, as summarized in Figure~\ref{fig:dara_service}. The first metric we examine is runtime. We find that Dara's runtime scales with the number of reference phases considered during the search (Figure~\ref{fig:dara_service}(a)). Since the uploaded patterns span a variety of chemical systems and the number of elements, runtimes vary accordingly. The median number of reference phases is 281 phases for one pattern, and the median runtime per pattern is 88.9 seconds, faster than a typical XRD data collection in a laboratory setting. Longer runtimes typically occur when the search space includes many reference phases, which can result from a chemical system with numerous elements or one that contains a large number of reference phases. In such cases, Dara may take several hours to complete an analysis. However, this can be significantly accelerated by deploying Dara on high-performance computing (HPC) clusters. With its parallel tree search implementation, Dara efficiently utilizes multiple cores/nodes to shorten processing times.
% Rwp
In terms of solution quality, Dara successfully identifies at least one solution for most patterns. Figure~\ref{fig:dara_service}(b) shows the distribution of $R_{wp}$. Dara achieves a median $R_{wp}$ of 5.85\%, with 78.1\% of patterns yielding values below 10\%. Although $R_{wp}$ alone does not fully determine fit quality \cite{toby2006r}, these statistics suggest that Dara performs robustly on real experimental lab data and provides a reliable starting point for human-guided phase identification and refinement.

Over several months of running this web application, we have observed some natural limitations that are also common to other XRD analysis software. The first is that the performance of Dara is highly related to the coverage of the experimental structure database in the chemical system of interest. For example, many structures have been well-characterized and deposited into structure databases within the \ce{Na2O}-\ce{Al2O3}-\ce{SiO2} chemical space. If one tries to analyze a pattern that contains Na, Al, Si, and O, it is more likely that Dara will find many solutions with various alternative phases. This is because the structures in this chemical space have been thoroughly studied, and numerous structures with minor structural modifications exist. On the other hand, if the sample contains phases that are not present in the structure database, Dara is likely to return either no result or phases that poorly fit the pattern. For instance, disordered rocksalt structures composed of Li-Mn-Ti-O have been extensively studied as promising cathode materials for next-generation lithium-ion batteries \cite{clement2020cation}. However, since no reference structure for these materials exists in the ICSD database, Dara fails to identify the Li-Mn-Ti-O phase when analyzing patterns from this composition space. 

For powder XRD analysis, especially when using a lab diffractometer, it is often impossible to fully resolve the structures. This can be attributed to the fact that lab-diffractometer-based powder XRD cannot provide sufficient information about composition and atomic positions, especially when the sample contains a mixture of multiple phases. To solve a pattern, it must be matched to known structures characterized by other methods. In recent years, the advancement of machine learning, particularly the emergence of generative models for crystal structures \cite{zeni2025generative, xie2021crystal}, has provided a viable approach to solving this problem. By conditioning the crystal generation on the fitness of XRD patterns and optionally the residual force in the lattice, one can obtain structures that can fit the XRD pattern while being physically meaningful. \cite{guo2025ab, riesel2024crystal, li2025powder} However, these models can only handle samples dominated by a single ordered phase, while real samples are often mixtures of known and unknown phases, potentially with disorder. One possible solution is to use phase analysis software, such as Dara, first to identify the known phases and then pass the unidentified peaks to crystal generation tools to search for a feasible structure resolution.

Another challenge is in assessing the chemical feasibility of specific phases. As Dara's tree search is designed to identify as many plausible phases as possible from the structural database, it often returns multiple phase combinations that fit the XRD pattern, some of which are unlikely to be present given the sample's synthesis conditions. For example, Dara may identify elemental metals such as lithium or sodium as having a good fit, even though these metals are highly reactive under ambient conditions where XRD measurements are typically performed. This occurs because elemental metals often have high-symmetry crystal structures, resulting in simple and strong diffraction peaks that can easily match those of nearby peaks in the measured pattern. While human experts often rely on this information to decide which phases to test for. However, automated XRD analysis tools typically do not incorporate knowledge of synthesis chemistry or stability, making it challenging to filter out chemically implausible phases without excluding valid ones. To address this limitation, information about the sample’s preparation conditions is essential. One possible improvement is to incorporate thermodynamic details into the analysis. For example, tools like reaction network analysis \cite{mcdermott2021graph} can be used to predict the likelihood of a phase forming under given reaction conditions. By considering the relative energies of reference phases, this approach can effectively eliminate thermodynamically unlikely phases from the analysis. Additionally, in recent years, large language models (LLMs) have emerged as promising tools for solving scientific problems in ways that mimic human reasoning \cite{lei2024materials, jablonka202314}. It is therefore possible to develop an LLM-based filter that leverages synthesis and chemistry knowledge to eliminate phases that are highly unlikely to exist in a given sample. This approach would enhance the chemical interpretability of XRD solutions by integrating domain knowledge into the automated analysis, as well as the information obtained from other characterization methods.

\section{Conclusion}
In this work, we present the design and performance evaluation of Dara, a tool for the automated analysis of powder XRD patterns. Leveraging a tree search algorithm for phase identification, a peak-matching algorithm for rapid identification of promising phases, robust full-profile fitting with the BGMN refinement engine, and an intelligent grouping algorithm for identified phases, Dara is capable of handling realistic powder diffraction samples with various sample effects and multi-phase mixtures. It is designed to address the ambiguity issue in XRD-based characterization and to provide reliable phase analysis by explicitly generating and testing alternative hypotheses with a refinement program, similar to how a human expert might analyze a powder sample of unknown phases. By comparing the performance of Dara with other analysis software as well as a human expert, we demonstrate that (i) Dara can match the performance of a human in analyzing the phase components of an XRD pattern, and (ii) it can efficiently analyze the pattern within a reasonable time with the full-profile fitting. As more tools like Dara are developed and integrated into autonomous synthesis workflows, we envision a future where high-throughput, expert-level structural analysis becomes routine, accelerating self-driving materials discovery and characterization at scale.

\section{Author Contributions}
Y.F. and M.J.M: conceptualization, software, writing - original draft, writing - review and editing. C.L.R. and S.W.: Validation and investigation. G.C.: resources, supervision, methodology, project administration, writing - review and editing.

\section{Conflicts of interest}
There are no conflicts to declare.

\section{Data availability}
The source code for Dara is available at \url{https://github.com/idocx/dara}. The version of Dara used in this study is \texttt{v1.0.0}. The benchmark summaries (Dara, Jade, and human) for the precursor mixture and pairwise reaction product dataset are available as two spreadsheets. Human analysis of the pairwise reaction product dataset is available in another spreadsheet. The raw XRD patterns used for this study are available at \url{https://zenodo.org/records/17410051}.

\section{Acknowledgement}
This work was primarily funded by the U.S. Department of Energy, Office of Science, Office of Basic Energy Sciences, Materials Sciences and Engineering Division under contract no. DE-AC02-05-CH11231 (D2S2 programme, KCD2S2). C.L.R was funded by the U.S. Department of Energy, Office of Science, Basic Energy Sciences, Division of Materials Science, through the Office of Science Funding Opportunity Announcement (FOA) Number DE-FOA-0002676: Chemical and Materials Sciences to Advance Clean-Energy Technologies and Transform Manufacturing. S.W. was funded by the U.S. Department of Energy, Office of Science, Office of Basic Energy Sciences, Materials Sciences and Engineering Division under contract No. DE-AC02-05-CH11231 (MINES project: The Science of Direct MINeral to Energy Storage Synthesis, FWP: FP00014914). S.W. was supported in part by the Jane Lewis Fellowship at UC Berkeley.

The authors thank Anubhav Jain (LBL), Lauren Walters (UC Berkeley), Bernardus Rendy (UC Berkeley), Tim Kodalle (LBNL), Guilhem Dezanneau (CNRS), Nobumichi Tamura (LBNL), Adam Corrao (BNL), Amalie Trewartha (TRI), Yan Zeng (FSU), and Olympia Dartsi (LBNL) for their testing and feedback. The authors would also like to express their special thanks to Nicola D\"{o}belin (RMS Foundation) and Reinhard Kleeberg (TU Bergakademie Freiberg) for guidance and permission on using BGMN refinement software as Dara's backend. Finally, we gratefully acknowledge our colleagues and peers for discussion and constructive feedback on the design of robust automated XRD analysis tools, which has greatly strengthened the development of this work.
\bibliography{ref}

\end{document}